\RequirePackage{fix-cm}

\documentclass[twocolumn]{svjour3}          
\smartqed  
\usepackage{graphicx}
\usepackage{amsmath,amsfonts,amssymb}
\usepackage{bm}
\usepackage{natbib}
\usepackage[colorlinks=true, allcolors=blue, breaklinks=true]{hyperref}
\usepackage{breakcites}
\newcommand{\todo}[1]{}
\renewcommand{\todo}[1]{{\color{red} TODO: {#1}}}%

\newcommand{\panelcase}{\uppercase}
\newcommand{\fig}[1]{Fig.~\ref{fig:#1}}
\newcommand{\figp}[1]{(Fig.~\ref{fig:#1})}
\newcommand{\panel}[2]{\fig{#1}\panelcase{#2}}
\newcommand{\panelp}[2]{(\panel{#1}{#2})}

\newcommand{\eq}[1]{equation~\eqref{eq:#1}}
\newcommand{\eqs}[2]{equations~\eqref{eq:#1}~and~\eqref{eq:#2}}
\newcommand{\eqrange}[2]{equations~\mbox{\eqref{eq:#1}--\eqref{eq:#2}}}

\begin{document}

\title{Cognitive swarming in complex environments with attractor dynamics and
  oscillatory computing
\thanks{This work was supported by NSF award NCS/FO 1835279, JHU/APL internal research and
development awards, and the JHU/Kavli Neuroscience Discovery Institute.}}

\titlerunning{Cognitive swarms with attractors and oscillators}  

\author{Joseph D. Monaco$^1$ \and
        Grace M. Hwang$^2$   \and
        Kevin M. Schultz$^2$ \and
        \mbox{Kechen Zhang$^1$}
}

\authorrunning{Monaco et al.} 

\institute{$^1$ Biomedical Engineering Department,
              Johns Hopkins University School of Medicine,
              Baltimore, MD, 21205, USA. \\
              \email{jmonaco@jhu.edu}  \\
           $^2$ The Johns Hopkins University/Applied Physics Laboratory,
              Laurel, MD, 20723, USA.
}

\date{Received: date / Accepted: date}

\maketitle

\begin{abstract}  
Neurobiological theories of spatial cognition developed with respect to
recording data from relatively small and/or simplistic environments compared to
animals' natural habitats. It has been unclear how to extend theoretical models
to large or complex spaces. Complementarily, in autonomous systems technology,
applications have been growing for distributed control methods that scale to
large numbers of low-footprint mobile platforms. Animals and many-robot groups
must solve common problems of navigating complex and uncertain environments.
Here, we introduce the `NeuroSwarms' control framework to investigate whether
adaptive, autonomous swarm control of minimal artificial agents can be achieved
by direct analogy to neural circuits of rodent spatial cognition.
NeuroSwarms analogizes agents to neurons and swarming groups to recurrent
networks. We implemented neuron-like agent interactions in which mutually
visible agents operate as if they were reciprocally-connected place cells
in an attractor network. We attributed a phase state to agents to enable
patterns of oscillatory synchronization similar to hippocampal models of
theta-rhythmic (5--12~Hz) sequence generation.
We demonstrate that multi-agent swarming and reward-approach dynamics can be
expressed as a mobile form of Hebbian learning and that NeuroSwarms supports
a single-entity paradigm that directly informs theoretical models of animal
cognition.
We present emergent behaviors including phase-organized rings and trajectory
sequences that interact with environmental cues and geometry in large,
fragmented mazes.
Thus, NeuroSwarms is a model artificial spatial system that integrates
autonomous control and theoretical neuroscience to potentially uncover common
principles to advance both domains.
\end{abstract}

\section{Introduction} \label{sec:introduction}

Spatial cognition in rodents has been extensively studied in non-naturalistic
environments such as linear or circular tracks, radial arm mazes, and T-mazes,
or small open-field arenas such as squares or cylinders of approximately
1--2~m$^2$ area. Such experimental conditions have allowed individual place
fields of hippocampal pyramidal neurons~\citep{OKeeDost71} and the activity of
other spatial cells~\citep{Knie06,MoseKrop08,SaveYoga08,PoulHart18,WangChen18}
to be exquisitely controlled and analyzed, leading to a
detailed neural coding account of distributed representations
that subserve spatial learning, memory, and planning in
mammals~\citep{OKeeNade78,MosePaul01,KnieHami11,MonaAbbo11,PfeiFost13,HartLeve14,
Burg14,SchiEich15,Fost17}, potentially extending to general cognitive
computations in humans~\citep{BellGard18,KunzMaid19}. However, the
multiplicity of Poisson-distributed hippocampal place fields exposed in larger
environments~\citep{FentKao08,RichLiaw14} and species differences in mapping
3-dimensional contexts~\citep{YartUlan13,CasaBush19} suggest large and/or
complex environments as the next frontier in understanding spatial navigation.

Computational models of rodent spatial networks have typically
emulated the restricted environments of experimental studies (for
computational efficiency, ease of analysis, and compatibility with
published data). Despite these limitations, recent theoretical results
have demonstrated the importance of sensory and cortical feedback in
stabilizing and shaping hippocampal and entorhinal cortical spatial
representations~\citep{MonaKnie11,PollNguy16,RennTort17,OckoHard18};
this relationship has been supported by experimental approaches
to the animals' own active sensing behaviors such as lateral head
scanning~\citep{MonaRao14,YadaDore17} and closed-loop control
of orienting distal cues~\citep{JayaMadh19}. Additionally,
extending prior theoretical frameworks such as the attractor map
formalism~\citep{Zhan96,Tsod99,SamsMcNa97,KnieZhan12} to large spatial
contexts has revealed substantial increases in the computational and
mnemonic capacities of these network models~\citep{HedrZhan16}. Thus, a
theoretical approach to naturalistic and dynamical spatial coding in large
or complex environments may require closed-loop systems that integrate
sensory information with internal spatial maps in continuously adapting
loops. Complementary to the animal studies, investigating the behaviors
and performance of completely specified but artificial spatial systems
including virtual agents and/or mobile robotics platforms may help to
elucidate the computational principles of spatial cognition in naturalistic
contexts~\citep{Hass18,TomoYaga18,SaveKnie19,GausBanq19}.

Hippocampal phenomena that have been theorized to support biological spatial
cognition include (1) self-stabilizing activity patterns in attractor map
networks and (2) temporal-phase organization relative to global oscillations. On
the basis of these phenomena, we introduce a brain-inspired dynamical controller
for self-organized swarms of autonomous agents. Our key realization was that
each virtual or robotic agent can be represented as a spatial neuron (e.g.,
a place cell) whose place preference follows the location of the agent in
its local environment. The analogy of a multi-agent group (or swarm) to a
space-coding neural network follows immediately. If we further suppose that
inter-agent distances map to `synaptic weights' and, consequently, that relative
agent movements map to changes in those weights, then spatial configurations of
the swarm constitute an attractor map network~\citep{Zhan96,Tsod99,SamsMcNa97}
and the swarm's internal motion dynamics constitute learning based on synaptic
modification~\citep{Hebb49,Oja82}.

Additionally, spatial activity in the rodent hippocampus is strongly
modulated by continuous theta oscillations~(5--12~Hz) during
locomotion~\citep{Vand69,Buzs05}. The resulting `phase precession,'
a monotonic advance in timing from late to early within each theta
cycle, may enhance the precision and temporal organization of
spatial codes~\citep{OKeeRecc93,JensLism00,FostWils07,DrieTodo18}
in ways that support decision-making and/or deliberative
planning during subsequent sleep or quiescent
states~\citep{Buzs89,JohnRedi07,BuzsMose13,WikeRedi15,PapaZiel16,MuesLase19}.
Recently, in contrast to phase precession, we discovered a novel class of
spatial phase-coding neurons in open field environments termed `phaser cells'
that were located predominantly in the rat lateral septum and characterized by
a strong coupling of theta-phase timing to firing rates~\citep{MonaDe-G19}.
This coupling supports an intrinsic neuronal mechanism of phase-coding that may
theoretically transform spatial information to synchronize downstream targets
using the global theta oscillation~\citep{MonaDe-G19}. Accordingly, to examine
the effects of temporal phase organization, our dynamical swarm controller
considers each agent to have an internal phase variable (analogous to the
theta-phase of a place cell or phaser cell) that may interact via oscillatory
coupling with its neighbors' phases. Such oscillator-based swarming has been
previously generalized as the `swarmalators' framework~\citep{OKeeHong17}.
Thus, together with attractor dynamics, these phenomena may provide brain-like
solutions to problems of decentralized self-organization and distributed
communication in autonomous swarming.

We refer to our conception of this swarm controller as
‘NeuroSwarms’~(\citealp{monaco2019cognitive}; \panel{neuroscience}{a}).
In this paper, we derive an operational NeuroSwarms
implementation~(Section~\ref{sec:methods}), present emergent
swarming behaviors in simulations of a fragmented and heterogeneous
environment~(Section~\ref{sec:emergent}), demonstrate NeuroSwarms as a dual
system which can be expressed through single-entity simulations that help
inform biological theory~(Section~\ref{sec:rewards}), evaluate adaptations of
reward-approach behaviors in a large hairpin maze~(Section~\ref{sec:large}),
and discuss implications for autonomous systems design and biological spatial
cognition in large, complex environments~(Section~\ref{sec:discussion}).

\begin{figure*}[!t]
  \includegraphics[width=1.0\textwidth]{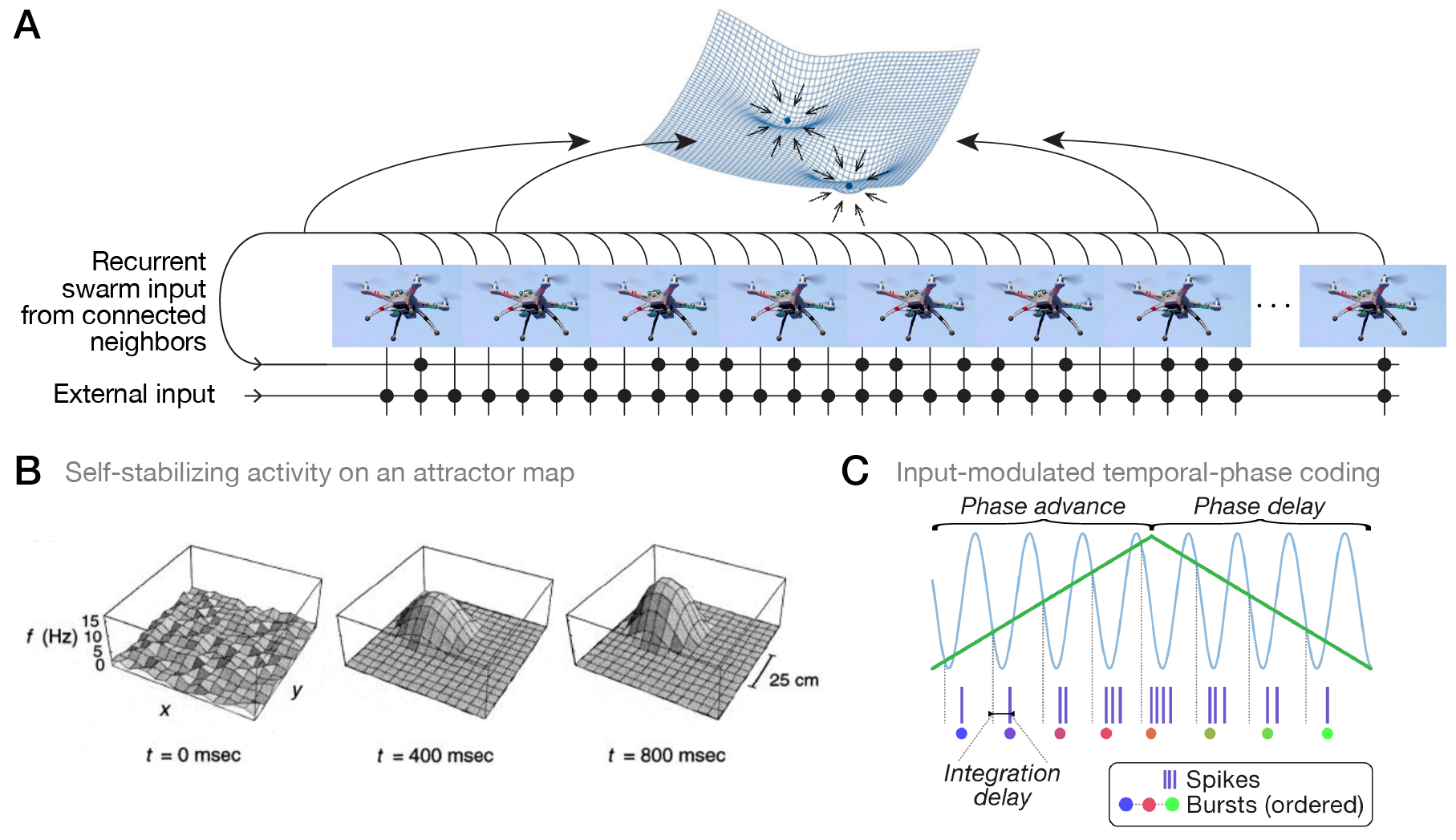}
  \caption{Conceptual schematic and theoretical neuroscientific inspiration for
  the NeuroSwarms controller. A. An artificial spatial system of mobile virtual
  or robotic agents communicate over sparse recurrent channels (bottom) just as
  spatial neurons in biological neural circuits produce reverberating activity
  patterns that reflect energy minima in the dynamical state-space of the system
  (e.g., fixed-point attractors; top; adapted from \citealp{KnieZhan12}). B.
  Example simulation of the spatial self-organization of an activity bump on an
  attractor map. In an attractor map network, the environment is represented by
  a continuum of locations with overlapping place fields, leading to network
  connectivity that produces self-reinforcing spatial activity patterns. Adapted
  from \citet{Zhan96}. C. Schematic of a minimal model of temporal-phase coding
  in which an excitatory external input (green) is rhthmically modulated by a
  continuous inhibitory oscillation (blue) such as the hippocampal theta rhythm.
  Adapted from \citet{MonaDe-G19} as permitted by the CC-BY 4.0 International
  License (creativecommons.org/licenses/by/4.0/).}
  \label{fig:neuroscience}
\end{figure*}

\section{Methods} \label{sec:methods}

\subsection{Hippocampal model mechanisms}

\paragraph{Self-stabilizing attractor maps.}
Hippocampal place cells fire within a contiguous region of the
animal's local environment, or `place field'~\citep{OKeeDost71}. Place
fields are thought to collectively form cognitive maps~\citep{OKeeNade78}
that are stabilized (at least in part) via attractor dynamics,
such as fixed points or continuous manifolds of the network energy
surface, that drive activity toward low-dimensional spatial or task
representations~(\panel{neuroscience}{b}; \citealp{KnieZhan12}). Attractor
map models have shown that recurrent connectivity between place cells with
nonlinear integration of inputs is nearly sufficient to achieve stable spatial
attractors~\citep{Zhan96,SamsMcNa97,Tsod99}. For instance, a rate-based network
following
\begin{equation}\label{eq:attractor}
    \frac{dr_i}{dt} = -r_i + g\left( \sum_j J_{ij}r_j + I_i \right) \, ,
\end{equation}
where $r_i$ is the rate of unit $i$, $I_i$ is the unit's total input, and $g$
is a sigmoidal nonlinearity, only further requires that the recurrent weights
$J_{ij}$ encode the degree of place-field overlap between units (i.e., the
strength of learned spatial associations). Such an encoding follows from a
kernel function of field-center distances, e.g.,
\begin{equation} \label{eq:kernel}
    J_{ij} := F(\mathbf{x}_i - \mathbf{x}_j) = A \
      \exp\left(-\frac{|\mathbf{x}_i-\mathbf{x}_j|^2}{\sigma^2}\right) - B \, ,
\end{equation}
where $\vec{x}_k$ is the field-center position of unit $k$, $\sigma$ is the
Gaussian scale constant, and $A$ and $B$ determine the strength of local
excitation vs. long-range inhibition, respectively. While this formulation
violates Dale's law, it illustrates the typical parsimony of attractor
map models~\citep{Tsod99}. A network constructed from \eqs{attractor}{kernel}
supports self-organization of its activity into a singular, contiguous `bump'
that emerges as the network relaxes (\panel{neuroscience}{b}; \citealp{Zhan96}).
The activity bump can then respond to input changes due to, e.g., movement
through the environment or internal processing. These conditions are
encapsulated within NeuroSwarms by analogizing (1) inter-agent visibility
(e.g., for a line-of-sight communication channel) to sparse, recurrent
network connectivity, and (2) a local kernel of inter-agent distances to
spatially-weighted synapses.

\paragraph{Oscillatory organization.}
Hippocampal phase precession relative to the theta rhythm is clearest on
linear tracks \citep{OKeeRecc93}, on which place fields typically arrange
in an unambiguous sequential order that may enable learning of temporally
compressed `theta sequences' \citep{FengSilv15}. While the phenomenon is less
clear in open, 2-dimensional environments, phase precession has been observed
in particular traversals of the firing fields of place cells as well as grid
cells in entorhinal cortex \citep{ClimNewm13,JeewBarr14}. However, our analysis
of phaser cells in the lateral septum \citep{MonaDe-G19} revealed a more direct
phase code for 2-dimensional position that was consistent with a minimal model
of temporal phase-coding~\panelp{neuroscience}{c}. To study how the phaser cell
code may contribute to sequence formation in open environments, the NeuroSwarms
controller adds an internal phase variable to each agent and couples the
modulation frequency of that phase to each agent's total input (see \eq{theta}
below).

\paragraph{Internalized place fields for neural control.}
There are two reasons why neural swarming control must decouple each
agent's physical location from its internal self-localization. First, the
multiplicity of agents is a qualitative difference with brain circuits;
every hippocampal neuron in a biological network corresponds to the same
single agent (e.g., a rat) and has particular inputs from internal processing
or sensory inputs~\citep{WangRoth16} that contribute to the appearance
and location of the cell's place field. Given the analogy of agents to
neurons~\panelp{neuroscience}{a}, an individual rat has many place fields
but an individual swarm agent should have only one. Further, the location of
an agent's place field cannot be identical to the agent's physical location,
which depends on the momentary vicissitudes of a entity operating in the
external world (or a simulation thereof). Second, experimental studies
have compellingly demonstrated that spatial path planning in hippocampal
networks may rely on activating sequences of place cells representing remote
locations~\citep{GuptMeer10,PfeiFost13,OlafBush18,MomeOtto18}, indicating that
internal representations should be separable from an animal's (or agent's)
current physical location. Thus, NeuroSwarms assigns a distinct cue-based place
preference to each agent.

\subsection{Mobile Hebbian learning with a global oscillation} \label{sec:learning}

Hebbian learning in neural network models typically increments or decrements
a synaptic weight according to a learning rate and a measure of the activity
correlation between the pre-synaptic (input) and the post-synaptic (output)
neurons~\citep{Hebb49,LevyStew79,Oja82,Eich18}. For the NeuroSwarms controller,
the conceptual similarity of the synaptic strength relation in a neural network
and the physical distance relation in a multi-agent group allows us to construct
a neural activation and learning model for the motion of artificial mobile
agents.

Following the Gaussian attractor-map kernel from \eq{kernel}, we explicitly
relate a recurrent synaptic weight matrix $\bm{W}\in\mathbb{R}^{N_s\times N_s}$,
prior to learning-based updates, to swarm state via
\begin{equation}\label{eq:Ws}
  W_{ij}=V_{ij}\exp(-D^2_{ij}/\sigma^2)\,,
\end{equation}
for inter-agent visibility $\bm{V}\in\{0,1\}^{N_s\times N_s}$, inter-agent
distances $\bm{D}$, and spatial constant $\sigma$. To provide for environmental
interactions, we consider a minimal reward-approach mechanism for a set of
reward coordinates that serve as attractive locations. Thus, we likewise
incorporate a feedforward matrix $\bm{W}^r\in\mathbb{R}^{N_s\times N_r}$ for
reward learning,
\begin{equation}\label{eq:Wr}
  W^r_{ik}=V_{ik}^r\exp(-D^r_{ik}/\kappa)\,,
\end{equation}
for agent-reward visibility $\bm{V}^r\in\{0,1\}^{N_s\times N_r}$, agent-reward
distances $\bm{D}^r$, and spatial constant $\kappa$. The reward weights are
based on an exponential kernel to allow for long-range approach behaviors. (We
emphasize that the NeuroSwarms framework encompasses the general equivalence
between synaptic weights and agent distances, but the particular implementation
that we present here is one of many possible designs.)

To define neuron-like inputs, we consider that each agent's internal
place-field location derives from the conjunction of sensory cue inputs related
to a preferred location. We define time-continuous sensory cue inputs
$\bm{c}\in\mathbb{R}^{N_s\times N_c}$ following
\begin{equation} \label{eq:cue}
  \tau_c\dot{c}_{ik} = V^c_{ik}V^{c^*}_{ik} - c_i \, ,
\end{equation}
for cue $k$, agent-cue visibility $\bm{V}^c\in\{0,1\}^{N_s\times N_c}$, fixed
agent-cue preferences $\bm{V}^{c^*}\in\{0,1\}^{N_s\times N_c}$, and integration
time-constant $\tau_c$. Thus, the product in \eq{cue} yields the integer
number of preferred cues visible from each agent's position. This means that
place-field size is not independently controlled but determined by the relative
cue richness of the environment: more cues will generally increase agent
heterogeneity and spatial selectivity. Similarly, we compute reward inputs
$\bm{r}\in\mathbb{R}^{N_s\times N_r}$ following
\begin{equation}
  \tau_r\dot{r}_{ik} = V^r_{ik}-r_i \,,
\end{equation}
for reward $k$, and integration time-constant $\tau_r$. Unlike sensory cues, all
agents respond equally to all (visible) rewards. Lastly, we define recurrent
swarming inputs $\bm{q}\in\mathbb{R}^{N_s\times N_s}$ following
\begin{equation} \label{eq:swarm}
    \tau_q\dot{q}_{ij} = V_{ij}\cos(\theta_j-\theta_i) - q_{ij} \, ,
\end{equation}
for `post-synaptic' agent $i$, `pre-synaptic' agent $j$, integration
time-constant $\tau_q$, and internal oscillatory phase state $\bm{\theta}$.
The cosine term in \eq{swarm} confers phase-modulation of the input in which
excitation/inhibition depends on whether units $i$ and $j$ are in/out of sync.

Having defined the input signals, we consider `net currents' as gain-modulated,
visibility-normalized quantities for sensory cue inputs,
\begin{equation} \label{eq:netcue}
  I_{c_i} = \frac{g_c}{\sum_{k}V^c_{ik}}\sum_{k=1}^{N_c}c_{ik} \,,
\end{equation}
reward inputs,
\begin{equation} \label{eq:netreward}
  I_{r_i} = \frac{g_r}{\sum_{k}V^r_{ik}}\sum_{k=1}^{N_r}W_{ik}^rr_{ik} \,,
\end{equation}
and recurrent swarming inputs,
\begin{equation} \label{eq:netrec}
  I_{q_i} = \frac{g_s}{\sum_{j}V_{ij}}\sum_{j=1}^{N_s}W_{ij}q_{ij} \,,
\end{equation}
where the parameter gains $g_c$, $g_r$, and $g_s$ sum to 1. Because the
net inputs are bounded in \eqrange{netcue}{netrec}, we simply apply linear
rectification rather than a saturating nonlinearity (cf.~\eq{attractor}) to
their sum to calculate post-synaptic activation
\begin{equation} \label{eq:activation}
  \bm{p} = \left[I_c+I_r+I_q\right]_+ \,,
\end{equation}
which is the remaining component needed to compute Hebbian (or any two-factor)
learning. In terms of swarming, however, the agents are phase-coupled via
\eq{swarm}. Thus, in the same way that a spiking neuron can be reduced to a
phase description of its orbit on the phase plane, we consider that $\bm{p}$
drives the agents' internal phase state, e.g.,
\begin{equation} \label{eq:theta}
  \dot{\bm{\theta}}=\bm{\omega}_0+\omega_I\,\bm{p}\,,
\end{equation}
where $\omega_I$ sets the maximum increase in input-modulated angular frequency
above the baseline frequency $\bm{\omega}_0$. The net effect of this mechanism
is that agents have place-cell-like spatial tuning with phaser-cell-like phase
coding and synchronization.

The core of the NeuroSwarms controller comprises the learning-based updates to
$\bm{W}$ and $\bm{W}^r$. A na\"{i}ve Hebbian rule, such as $dW_{ij}=\eta p_i
q_j$, would cause weights to grow unbounded, leading to ictogenesis in recurrent
networks or spatial singularities in swarms. Instead, after updating agent
activations via \eq{activation}, we compute updated weights $\bm{W}^\prime$ as
\begin{equation} \label{eq:oja}
  W^{\prime}_{ij} = W_{ij}+\mathrm{\Delta}t\,\eta V_{ij}\,p_i(q_{ij}-p_iW_{ij}) \,,
\end{equation}
with simulation time-step $\mathrm{\Delta}t$ and learning rate $\eta$, which
effectuates a pre-synaptic normalization according to Oja's rule~\citep{Oja82}.
Similarly, the updated feedforward weights $\bm{W}^{r\prime}$ are computed for
reward $k$ as
\begin{equation} \label{eq:oja_r}
  W^{r\,\prime}_{ik}=W^r_{ik}+\mathrm{\Delta}t\,\eta_r V^r_{ik}\,p_i(r_{ik}- \
    p_iW^r_{ik}) \,.
\end{equation}
The normalization effected by \eqs{oja}{oja_r} is due to a subtractive term,
quadratic in the post-synaptic activation $\bm{p}$, that depresses the growth
of overly active synapses. In place-cell network models, feedback inhibition
typically serves to spread out place fields to more efficiently map an
environment~\citep{SaveKnie10,MonaAbbo11}, but the lack of explicit inhibition
in NeuroSwarms allows synaptic depression to provide a similar repulsive role
due to the distance--weight equivalence of \eq{Ws}.

\subsection{Neural swarm control: closing the loop} 
\begin{figure*}[!tb]
  \centering
  \includegraphics[width=\textwidth]{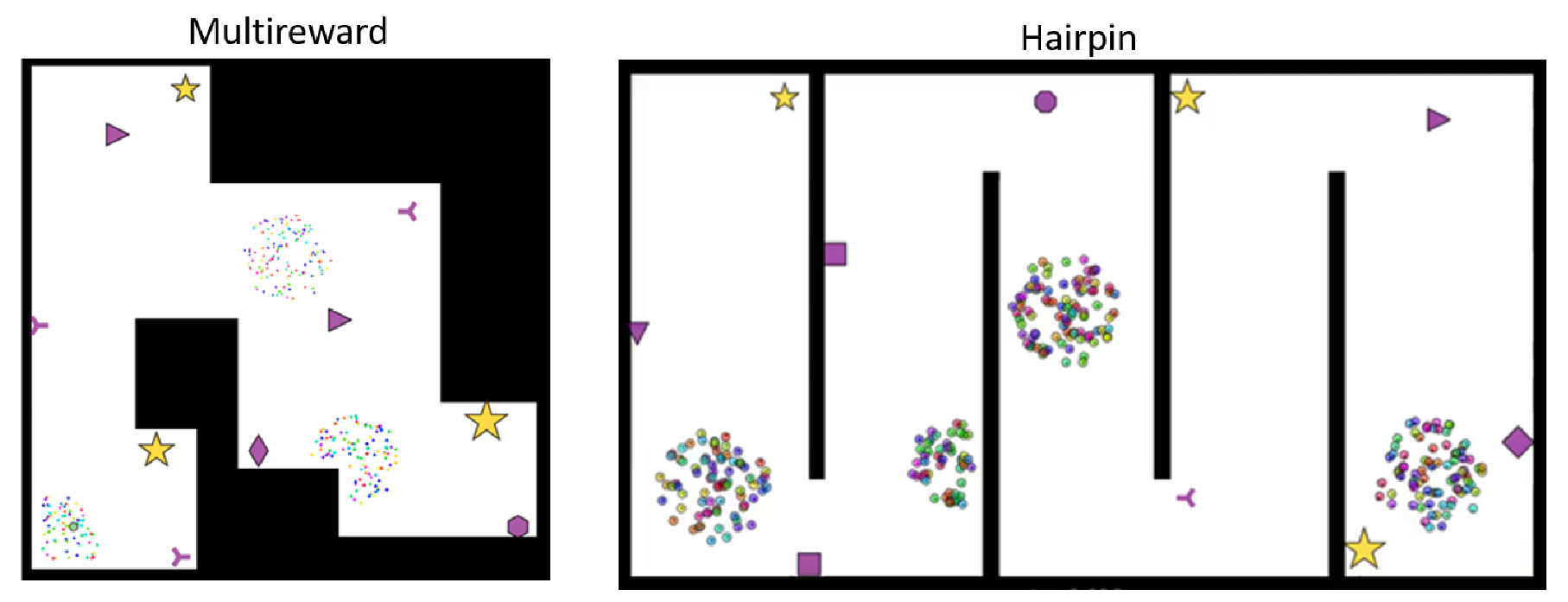}
  \caption{Example post-initialization ($t=0.1$~s) swarm states for NeuroSwarms
  simulations. (Left) A single-agent simulation in the `multireward' arena,
  which contains 3~rewards (gold stars; northwest, southwest, southeast), 7~cues
  (purple shapes), and 3~spawn discs. White enclosed areas constitute the set
  of allowable locations for swarm agents; black regions constitute barriers
  and disallowed locations. Initial particle positions are sampled from the
  spawn discs and initial phases are random. Green circle in southwest: the
  single agent; dots: 300 virtual swarm particles with internal phase state
  indicated by color. (Right) A multi-agent simulation in the `hairpin' maze,
  which contains 5~connected hallways, 3~rewards, 7~cues, and 4~spawn discs.
  Circles: 300~swarm agents with internal phase state indicated by color; reward
  (gold star) size is for visual differentiation only and has no effect in the
  model.}
  \label{fig:mazes}
\end{figure*}

To integrate with swarming, the controller attempts to drive the agents'
kinematic states to the equivalent desired inter-agent distances, in effect
replacing the typical attraction and repulsion fields of conventional swarming
models~\citep[e.g.,][]{gazi2011swarm}. The updated weights $\bm{W}^{\prime}$ and
$\bm{W}^{r\,\prime}$ can be converted directly to desired distances by inverting
the Gaussian swarming kernel in \eq{Ws},
\begin{equation}
  D^\prime_{ij} = \sqrt{-2\sigma^2\log W^\prime_{ij}} \,,
\end{equation}
and the exponential reward kernel in \eq{Wr},
\begin{equation}
  D^{r\,\prime}_{ij} = -\kappa\log W^{r\,\prime}_{ij} \,.
\end{equation}
To compute the resultant swarm motion, the desired positional shift of agent $i$
is averaged across its visible neighbors, i.e.,
\begin{equation}
    \bm{f}_i = \frac{1}{2\sum_{j} V_{ij}} \sum_{j=1}^{N_s}V_{ij}\,(D^\prime_{ij}
      \ - D_{ij})\frac{\vec{x}_j-\vec{x}_i}{|\vec{x}_j-\vec{x}_i|} \,,
\end{equation}
and the resultant reward-related motion is similarly computed as the average
\begin{equation}
    \bm{f}^r_i = \frac{1}{\sum_k V^r_{ik}} \sum_{k=1}^{N_r}
      V^r_{ik}\,(D^{r^\prime}_{ik} \ -
        D^r_{ik})\frac{\vec{x}^r_k-\vec{x}_i}{|\vec{x}^r_k-\vec{x}_i|} \,.
\end{equation}
The net positional shift is calculated as a linear combination of the swarm-
and reward-related shifts,
\begin{equation}
  \mathrm{\Delta}{\vec{x}} = \alpha \bm{f}+(1-\alpha)\bm{f}^r\,,
\end{equation}
where $\alpha=0.5$ for all simulations presented. The remaining processing of
$\mathrm{\Delta}{\vec{x}}$ in our NeuroSwarms implementation serves to embed
the foregoing dynamics within `physical' simulations of irregular or complex
2-dimensional environments. First, our example evironments~\figp{mazes} of
\ensuremath{\sim}500-point height (for arbitrary points units) were processed
for wall proximity and normal vectors for all interior locations. Thus, a
`barrier aware' positional shift $\mathrm{\Delta}\vec{x}^b$ is calculated as
\begin{equation} \label{eq:barrier}
  \mathrm{\Delta}\vec{x}^b_i = (1-\beta_{s_i})\mathrm{\Delta}\vec{x}_i + \
    \beta_{s_i}|\mathrm{\Delta}\vec{x}_i|\,\vec{n}_{s_i} \,,
\end{equation}
for an exponential kernel $\bm{\beta}_s=\exp(-\bm{d}/\lambda)$ of
\mbox{distance $\bm{d}$} to the nearest wall with a constant of
$\lambda=20$~points, and the normal vectors $\vec{n}_s$ of the
nearest wall. These shifts update the internal place-field locations
$\vec{x}_s\leftarrow\vec{x}_s+\mathrm{\Delta}\vec{x}^b$ of each swarm agent.
Second, `physical' agent locations are updated based on the instantaneous
velocity needed for each agent to approach their internal field locations,
$\vec{v}_s = (\vec{x}_s-\vec{x}) / \mathrm{\Delta}t$, which is processed through a
momentum filter,
\begin{equation} \label{eq:momentum}
  \vec{v}_\mu=\mu\vec{v}+(1-\mu)\vec{v}_s \,,
\end{equation}
with the actual velocity (prior to updating) $\vec{v}$ and coefficient
$\mu$; a speed-limiting nonlinearity based on a kinetic-energy maximum
$E_{\mathrm{max}}$,
\begin{align}
\label{eq:agent_dynamics}
    &\vec{v}_{\mathrm{max}} = \sqrt{2E_{\mathrm{max}}/\bm{m}} \,, \\
    &\vec{v}_{k_i} = \vec{v}_{\mathrm{max}_i}\tanh\left(\frac{|\vec{v}_{\mu_i}|}{ \
      \vec{v}_{\mathrm{max}_i}}\right) \!\!\frac{\vec{v}_{\mu_i}}{|\vec{v}_{\mu_i}|} \,,
\end{align}
where $\bm{m}$ is agent mass; and barrier avoidance as in \eq{barrier},
\begin{equation} \label{eq:vwalls}
  \vec{v}_i = (1-\beta_i)\vec{v}_{k_i} + \beta_i|\vec{v}_{k_i}|\,\vec{n}_i \,,
\end{equation}
for proximity $\bm{\beta}$ and normal vectors $\vec{n}$. Finally, agent
locations are updated by $\vec{x}\leftarrow\vec{x}+\vec{v}\mathrm{\Delta}t$.

\subsubsection{Single-entity simulations} \label{sec:singlemethods}
In keeping with the neuroscientific motivation for
NeuroSwarms~\figp{neuroscience}, our implementation allows for singleton
simulations analogous to conventional models of neural networks in an animal
such as a rat navigating a maze. With only minor adjustments, NeuroSwarms can
operate with a single agent (i.e., $N=1$) that owns a collection of `virtual'
or `mental' swarming particles (e.g., $N_s=300$) that guide the agent's spatial
behavior. In this sense, the virtual swarm represents a highly dynamic spatial
field that provides the agent with various options for constructing a path
through the environment. The dynamics of the virtual swarm are as described
above up to \eq{barrier}. An array $\bm{V}^\delta\in\{0,1\}^{N_s}$
indicates which particles' positions are visible to the agent and serves to
additionally mask the learning updates in \eqs{oja}{oja_r}. To produce motion,
single-agent velocity is instead calculated using a cubic-activation-weighted
average of the swarm,
\begin{equation} \label{eq:vsingle}
  \vec{v}_s = \frac1{\mathrm{\Delta}t\sum_jV_j^\delta p_j^3}\sum_{i=1}^{N_s}V^\delta_i \
    p_i^3 (\vec{x}_{s_i}-\vec{x}) \,,
\end{equation}
prior to processing the `physical' embedding of the agent's motion in
\eqrange{momentum}{vwalls}. Thus, the agent constructs a path toward the most
highly activated and visible swarm particles.

\subsection{NeuroSwarms simulations}
Simulated environments \figp{mazes} contain fixed-position rewards and cues
depicted as gold stars and purple shapes, respectively. Environments are
otherwise defined by a set of linear barrier segments (e.g., walls) that form
a closed shape defining an interior space that becomes the set of allowable
agent locations. Simulations are initialized by setting all velocities, input
signals, and activations to zero, randomly choosing internal phase states,
and randomly assigning agent positions to allowable locations within a set of
`spawn discs' defined in the environment. Random number generator seeds are
reused for simulations meant to compare parameter values, unless otherwise
specified. Environments are specified as vector image files in Tiny SVG format
with XML text nodes defining reward, cue, and spawn disc locations. Unless
noted, parameters were set to the default values displayed in Table~\ref{tab:1}.
The python source code will be made available upon reasonable request.

\begin{table}[!hb]
  \caption{Parameters, default values, and descriptions (with units) for the
  NeuroSwarms controller implementation.}
  \label{tab:1}
  \begin{tabular}{lll}
    \hline\noalign{\smallskip}
    $\mathrm{\Delta}t$    & 0.01  & s, integration time-step of simulation \\
    duration              & 180.0 & s, total simulation time \\
    $N$                   & 300   & no. of `physical' agents (multi-agent)\\
    $N$                   & 1     & no. of `physical' agents (single-entity) \\
    $N_s$                 & 300   & no. of internal fields (multi-agent) or \\
                          &       & virtual particles (single-entity)\\
    $D_{\rm max}$$^\dag$  & 1.0   & max. inter-agent visibility range \\
    $E_{\rm max}$         & 3e3   & kg$\cdot$points$^2$/s$^2$, max. kinetic energy \\
    $\mu$                 & 0.9   & momentum coefficient of agent motion \\
    $m_{\rm multi}$       & 0.3   & kg, mean agent mass (multi-agent) \\
    $m_{\rm single}$      & 3.0   & kg, agent mass (single-entity) \\
    $\sigma$$^\dag$       & 1.0   & spatial scale of swarm interaction \\
    $\kappa$$^\dag$       & 1.0   & spatial scale of reward interaction \\
    $\eta$                & 1.0   & learning rate for swarm connections \\
    $\eta_r$              & 1.0   & learning rate for reward connections \\
    $\omega_0$            & 0.0   & cycles/s, baseline oscillatory frequency \\
    $\omega_I$            & 1.0   & cycles/s, max. increase in oscillatory \\
                          &       & frequency due to neural activation\\
    $g_c$                 & 0.4   & gain of sensory cue inputs \\
    $g_r$                 & 0.2   & gain of reward inputs \\
    $g_s$                 & 0.4   & gain of swarming inputs \\
    $\tau_c$              & 0.5   & s, time-constant of sensory cue inputs \\
    $\tau_r$              & 0.5   & s, time-constant of reward inputs \\
    $\tau_q$              & 0.1   & s, time-constant of swarming inputs \\
    $d_{\rm rad}$         & 0.0   & points, reward contact radius \\
    \noalign{\smallskip}
    \hline
    \noalign{\smallskip}
  \end{tabular}
  \dag~These parameter values are multiplicatively scaled to the notional
  environment size, defined in points as the radius of a disc with the same area
  as the set of allowable locations in the environment's interior.
\end{table}

\section{Results} \label{sec:results}

\subsection{Emergent swarming behaviors} \label{sec:emergent}

\begin{figure*}[!tb]
  \includegraphics[width=1\textwidth]{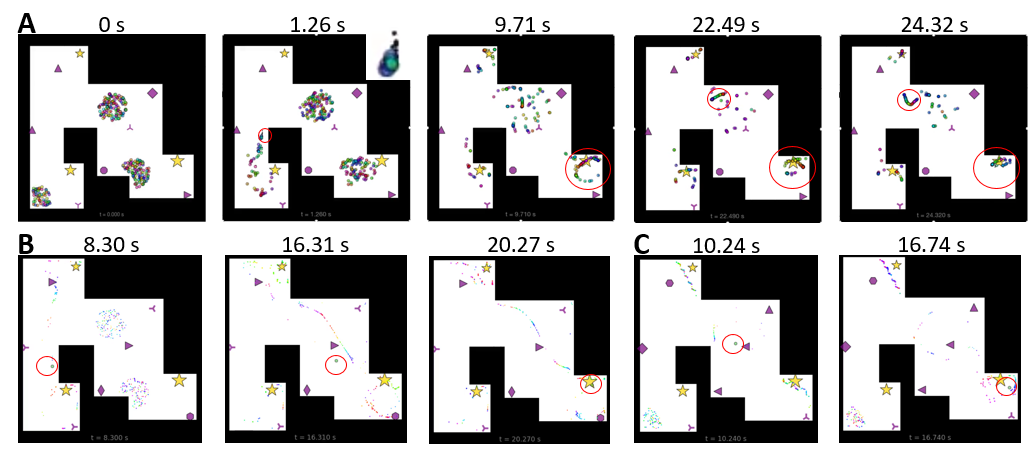}
  \caption{Temporal evolution of swarming and single-entity approaches to
  rewards. A. Three agent-clusters were initially populated in the multireward
  arena. The internal place-field location of each agent is indicated by a
  small black dot (e.g., $t=1.26$~s, inset, top right). Phase sorting is
  indicated by sequentially ordered colors of the circle markers representing
  agent positions. A reward-centered phase ring was created ($t=9.71$~s)
  with a decreasing diameter over time (southeast corner, $t=22.49$~s and
  $t=24.32$~s). NeuroSwarms parameters: $\sigma=1.5$, $g_c = 0.2$, $g_r=0.3$,
  $g_s=0.5$; Table~\ref{tab:1}.
  B. A single-entity agent (green circle) was guided by $N_s=300$ virtual
  particles (phase-colored dots). Swarm particles formed phase sequences
  leading the agent from the southwest corner to the reward location
  in the southeast corner of the arena by $t=20.3$~s. NeuroSwarms
  parameters: $\sigma=4$, $\kappa=1.5$, $g_c = 0.2$, $g_r=0.3$, $g_s=0.5$;
  Table~\ref{tab:1}.
  C. Step-like patterns of particles appeared near rewards that were occluded
  from the perspective of the single agent by corners in the environmental
  geometry. While the agent became `indecisive' around $t=10.24$~s, as it was
  pulled simultaneously in both directions, the agent ultimately found its
  way to the southeast reward by $t=16.74$~s. NeuroSwarms parameters: 
  $\sigma=4$, $\kappa=8$, $g_c=0.2$, $g_r=0.3$, $g_s=0.5$; Table~\ref{tab:1}.}
  \label{fig:emergence}
\end{figure*}

We designed the multireward arena~(\fig{mazes}, left) to characterize emergent
swarming and reward approach behaviors, and the hairpin maze~(\fig{mazes},
right) to assess behavioral adaptation in large, fragmented environments.
We observed several emergent dynamical behaviors in simulations of both
multi-agent swarming and single-entity locomotion~(Section~\ref{sec:methods},
Methods). The most notable and persistent behaviors included the emergence of
phase-sorted spatial formations such as line segments, rings, or concentric
loops~\figp{emergence}. These behaviors were analogous in form to (1) the
`phase wave' states observed in certain swarmalator regimes~\citep{OKeeHong17},
and (2) the hippocampal phenomena of theta sequences and theta-rhythmic phase
assemblies~\citep{FostWils07,DrieTodo18}. Further, by inspection of simulation
movies, we observed two dynamical features. First, agent subgroups forming line
segments and rings continuously phase-synchronized in a shared oscillation
that was independent from the absolute movement or rotation of the formation
in space. Second, line or ring formations would often break apart and re-form
new configurations that typically involved other agents or formations
that were able to phase-synchronize with elements of the subgroup. These
alternating disintegrative and aggregative dynamics may be consistent with
analyses of persistent homologies in place-cell networks with transient
connectivity~\citep{BabiDaba17}.

These spatiotemporal dynamics are evident across frame captures of
multi-agent~\panelp{emergence}{a} and single-entity~\panelp{emergence}{b}
simulations. While phase-ordered groups could appear far from
rewards~(\panel{emergence}{A}, last two frames, smaller red circles), swarm
agents typically approached a reward location and formed a rotating ring
centered on the reward position~(\panel{emergence}{A}, southeast corner, last
three frames). Such reward rings appeared in single-entity simulations, but the
virtual swarm particles~(Section~\ref{sec:singlemethods}) additionally exhibited
particularly extended line segments that often traced out phase-ordered
trajectory sequences; e.g., the agent followed an extended sequence to the
reward located in the southeast corner~(\panel{emergence}{b}, last two frames).
Further, we observed that the size of reward rings decreased over time,
reflecting a relaxation of phase and momentum given the centrally organizing
reward location.

When the reward kernel's spatial scale $\kappa$ (\eq{Wr}; Table~\ref{tab:1})
was increased, streams of virtual swarm particles formed around distal rewards
as the particles' motion was modulated by agent visibility interacting
with the geometry of the environment. As shown in the first frame of
\panel{emergence}{C}, a step-like pattern formed near the northwest
reward location while a wavy pattern formed near the southeast reward
location. Both virtual swarm formations presented path choices to the
single agent located in the large central compartment of the arena. As
expected~(Section~\ref{sec:singlemethods}), virtual swarm particles that were
not visible to the agent remained fixed in place due to masking of the weight
updates in \eqs{oja}{oja_r}. In addition to single rings, double and even
triple concentric loops of nested, non-overlapping, phase-sorted rings were
observed in some simulations. An example of a double loop forming is shown
in the southeast corner at $t=16.74$~s~\panelp{emergence}{C}. Strikingly,
we did not predict, expect, or adapt the NeuroSwarms controller design
to observe these emergent behaviors; we simply implemented abstractions
for place-cell spatial tuning, phaser-cell oscillatory synchronization,
and a distance--weight equivalence for Hebbian learning with notions of
visibility and environmental geometry that provided spatial barriers to
communication~(Section~\ref{sec:methods}). These behaviors would be unexpected
as well from conventional swarming algorithms~\citep{gazi2011swarm}.

\subsection{Reward-based behavior in a compartmented arena} \label{sec:rewards}

To assess the spatial performance of NeuroSwarms, we examined the ability
of single-entity behavior to find all three rewards in the multireward
arena. We focused on the parameter constants governing the spatial scale
of swarm ($\sigma$) and reward ($\kappa$) interactions (\eqs{Ws}{Wr};
Table~\ref{tab:1}) and found ($\sigma$, $\kappa$) values for which the
agent approached multiple rewards regardless of its initial spawn location.
Due to the random initialization of location, we selected 40 simulations
for analysis in which the agent was spawned in the southwest corner~(as in
\fig{mazes}, left). The agent successfully captured one, two, or all three
rewards in 11, 28, and 1 simulation(s) at elapsed times ranging from 4--108,
20--179, and \ensuremath{\sim}160~s, respectively. Frame captures of reward
approaches are shown in \panel{rewardseeking}{a} for the simulation in which
all three rewards were found. The ability of the agent to approach multiple
fixed rewards over time was an unexpected and emergent behavior: based on our
NeuroSwarms implementation, we had predicted that the rewards would serve as
stable attractors in the absence of additional mechanisms such as adaptation or
reward learning. However, in accordance with those expectations, we observed
simulations which failed to explore much of the environment after approaching
a single reward location. For the same parameters but a different random
seed than shown in \panel{rewardseeking}{a}, a failed exploration occurred
\panelp{rewardseeking}{b} when the virtual particles split into two fixed-point,
out-of-phase attractors that essentially `trapped' the agent.

To counteract these unsuccessful equilibria, we implemented a `reward
capture' mechanism in the NeuroSwarms controller based on a minimum contact
radius, $d_{\rm rad}$. This feature causes rewards to cease being attractive
locations to the virtual swarm particles upon contact by the agent, thus
releasing the agent from reward-related attractors before further exploration
is prevented. Indeed, having capturable rewards with $d_{\rm rad}=12$~points
enabled a simulation that was otherwise identical to the failed case
\panelp{rewardseeking}{b} to successfully navigate the arena to capture all
three rewards \panelp{rewardseeking}{c}. Thus, a notion of reward adaptation or
reward consumption may be crucial to achieving continuous exploration.

\begin{figure*}[tb]
  \includegraphics[width=1\textwidth]{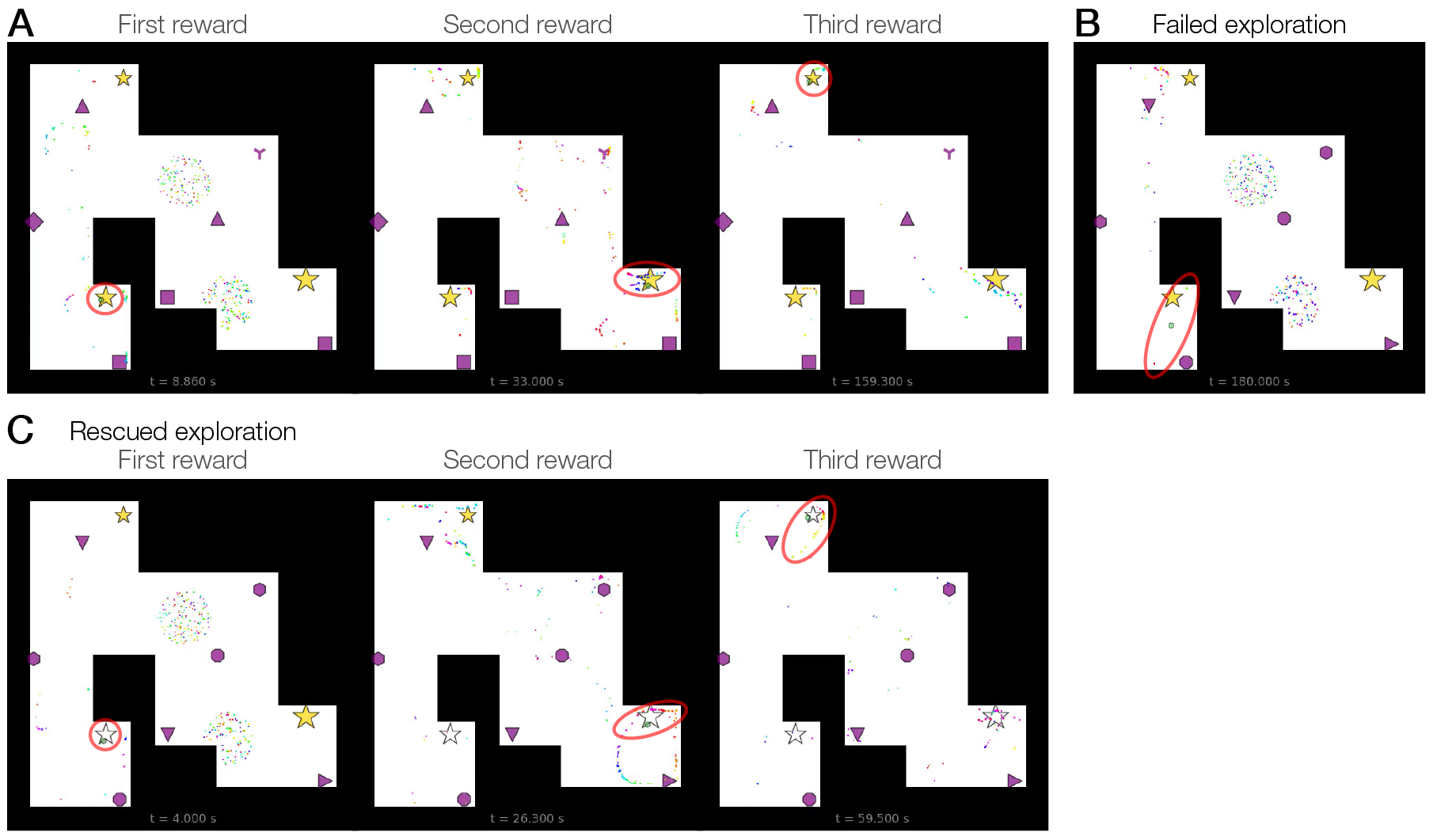}
  \caption{Single-entity reward-approach behavior with fixed or capturable
  rewards. The agent was initialized to the spawn disc in the south west corner of
  the multireward arena.
  A. A rare example in which the single agent (green circle) captured all three
  rewards when rewards were fixed (i.e., they remained attractive despite previous
  contact with the agent): southwest reward at \ensuremath{\sim}8.9~s, southeast
  reward at \ensuremath{\sim}33~s, and northwest reward at \ensuremath{\sim}160~s.
  Movie frames show the initial contacts with each reward (gold stars).
  NeuroSwarms parameters: $\sigma=4$, $\kappa=1.5$, $g_c = 0.2$, $g_r=0.3$,
  $g_s=0.5$; Table~\ref{tab:1}.
  B. With the same parameters as (A) but initialized with a different random
  seed, this final frame of a simulation shows the converged state after the agent
  was attracted to the southwest corner and remained there for the duration.
  The red ellipse highlights that the agent was `stuck' between two fixed-point
  attractors that formed through mutual phase-desynchronization.
  C. With the identical parameters and random seed as (B), rewards were made
  to be `capturable' at a minimum contact radius of $d_{\rm rad}=12$~points.
  Thus, rewards ceased to be attractive locations once the agent made initial
  contact. The agent captured the southwest reward at \ensuremath{\sim}5~s,
  the southeast reward at \ensuremath{\sim}27~s, and the northwest reward at
  \ensuremath{\sim}60~s. Transparent/white stars indicate captured rewards.
  NeuroSwarms parameters: $\sigma=4$, $\kappa=1.5$, $g_c=0.2$, $g_r=0.3$,
  $g_s=0.5$, $d_{\rm rad}=12$; Table~\ref{tab:1}.}
  \label{fig:rewardseeking}
\end{figure*}

For the 40 single-entity simulations with fixed rewards, the bottom panel
of \panel{contactradius}{A} reveals strong attractors at the southeast and
northwest corners of the arena associated with reward locations. To demonstrate
the effect of the contact radius on exploration when rewards were capturable,
the trajectories resulting from contact radii of 1, 4, 10, and 15~points are
shown in the top row of \panel{contactradius}{A}; these values produced 1, 3,
8, and 30 (out of 40) simulated trajectories, respectively, that successfully
contacted all three rewards (\panel{contactradius}{A}, red traces). In a few
simulations, the single-entity agent spawned in the southwest corner, found
the southeast reward first, and then later returned to the southwest corner in
order to collect all three rewards; such a wandering trajectory suggests that
the model might qualify as an ergodic system under these conditions, but that
hypothesis would be appropriately addressed by future analytical studies. These
results demonstrate that the sensitivity of reward capture modulates exploratory
variability by mitigating the effect of reward-related attractors. Histograms
of the time-to-capture profile across agent spawn sites and reward locations
reflect the structure of the environment as well as the different possible
sequences of reward contact \panelp{contactradius}{b}. Thus, the contact radius
for capturable rewards exerted substantial control over the likelihood of the
single-entity agent finding all rewards in the environment.

\begin{figure*}[tb]
  \includegraphics[width=1\textwidth]{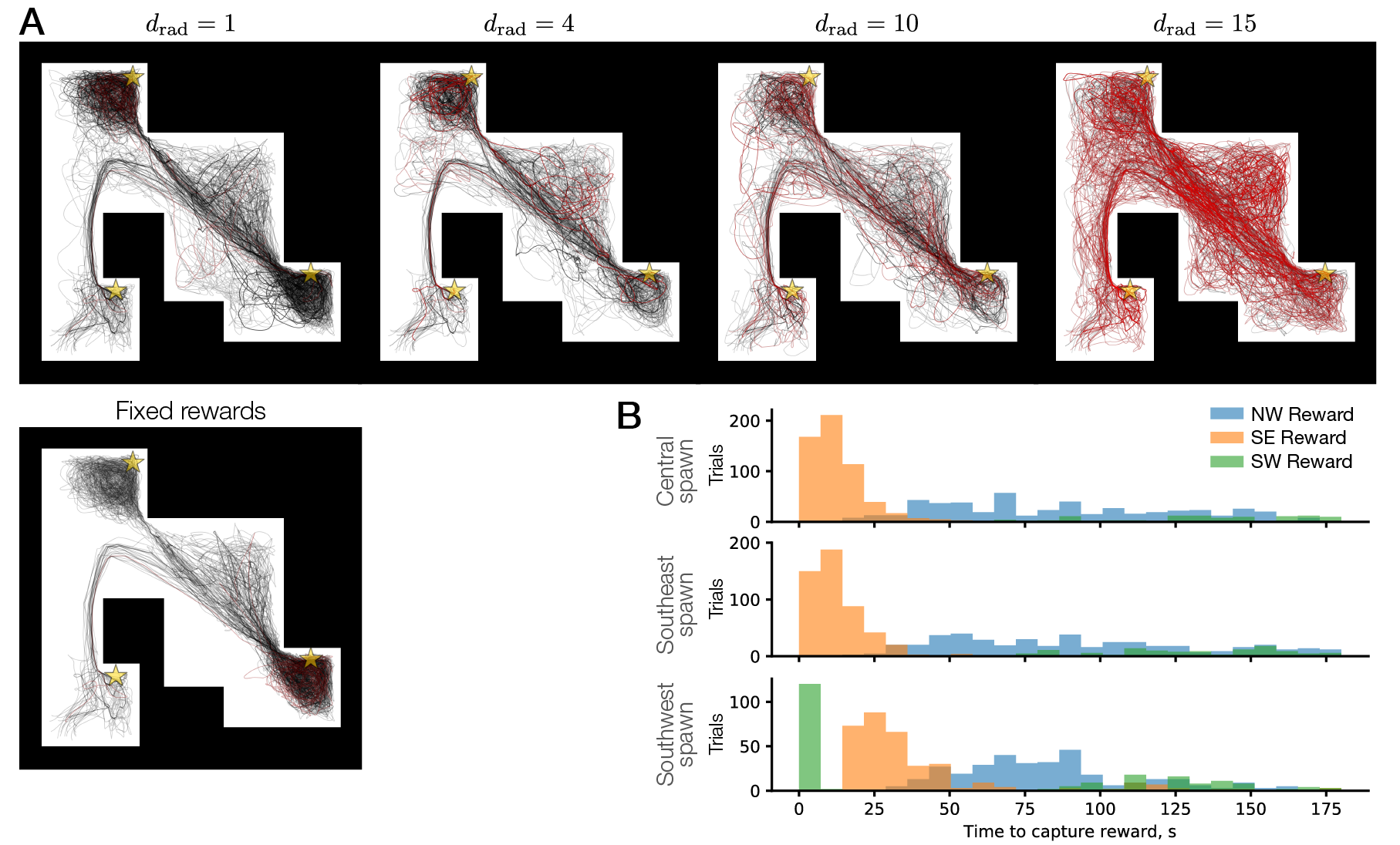}
  \caption{Dispersion of exploratory trajectories with capturable rewards.
  A. Superimposed agent trajectories are shown from 40 single-entity simulations
  of 180~s duration in which the agent was initialized to the southwest
  corner~(Section~\ref{sec:singlemethods}).
  With fixed (non-capturable) rewards, only 1 simulation (bottom, red trace)
  contacted all three rewards in the arena (see \panel{rewardseeking}{a}) and
  there was minimal variance in the exploratory paths taken by the agent in
  the other simulations (black traces). The dense sampling of the northwest
  and southeast reward location indicates these were strong attractors for the
  agent.
  With increasing contact radii of 1, 4, 10, or 15~points~(top), exploratory
  variance increased, the reward attractors became relatively weaker, and higher
  proportions of agent trajectories successfully visited all three rewards
  (red traces). NeuroSwarms parameters: $\sigma=4$, $\kappa=1.5$, $g_c=0.2$,
  $g_r=0.3$, $g_s=0.5$. Gold stars: reward locations.
  B. For 700 single-entity simulations with random initial agent locations and
  $d_{\rm rad}=15$, histograms for each of the agent spawn locations (central,
  southeast, or southwest) display the time-to-capture profile of each of the
  three rewards. NeuroSwarms parameters same as the top right panel of (A).}
  \label{fig:contactradius}
\end{figure*}

\subsection{Behavioral adaptation in large hairpin mazes} \label{sec:large}

A key challenge facing current state-of-the-art swarm controllers is the
inability to rapidly adapt to dynamic changes in complex environments. The
hairpin maze is well suited to study such adapation because swarm agents spawned
from certain hallways do not have line-of-sight visibility to rewards that may
be located in adjacent hallways. A form of behavioral adapation can be assessed
based on whether agents spawned into reward-free hallways can nonetheless
navigate to rewards in other parts of the maze.

\begin{figure*}[tb]
  \includegraphics[width=1\textwidth]{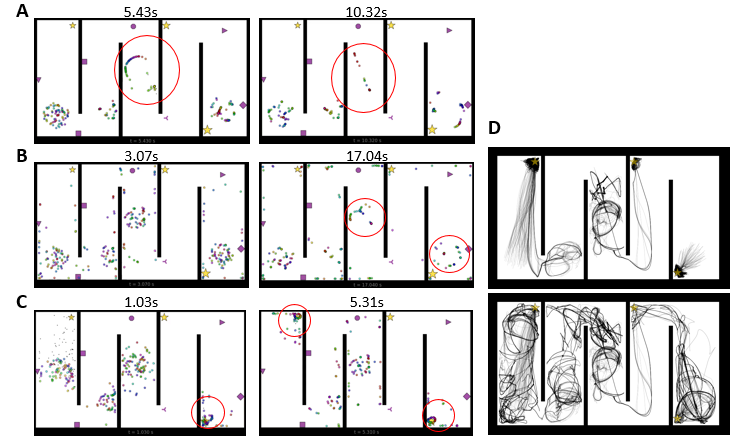}
  \caption{Dynamics of a multi-agent swarm in a large hairpin maze. Movie frame
  captures are shown for simulations with $N=300$ agents in a rectangular
  environment ($885\times 519$ points including borders) partitioned into 5
  hallways in a hairpin pattern. Three hallways contain rewards which are
  substantially occluded from the other maze sections. Emergent formations are
  circled in red.
  A. Frames from a pure swarming simulation, without reward or sensory cue
  influence. NeuroSwarms parameters: $d_{\rm max}=1.5$, $\eta=1$, $\eta_r=0$,
  $g_c=0$, $g_r=0$, $g_s=1$; Table~\ref{tab:1}.
  B. Frames from a capturable-rewards simulation with 1:1 swarm/cue input gains
  but no reward influence. NeuroSwarms parameters: $d_{\rm max}=1.5$, $\eta=1$,
  $\eta_r=0$, $g_c=0.5$, $g_r=0$, $g_s=0.5$; Table~\ref{tab:1}.
  C. Frames from a capturable-rewards simulation with equalized swarm, reward,
  and cue input gains. NeuroSwarms parameters: $d_{\rm max}=1.5$, $\eta_s=1$,
  $\eta_r=1$, $g_c=g_r=g_s=1/3$; Table~\ref{tab:1}.
  D. Multi-agent trajectories are shown from two 80~s simulations: fixed
  rewards (top) and capturable rewards with $d_{\rm rad}=10$~points (bottom).
  Compare with multireward arena simulations in \panel{contactradius}{a}.
  NeuroSwarms parameters: $d_{\rm max}=1.5$, $\sigma=2$, $\kappa=6.6$,
  $g_c=0.1$, $g_r=0.1$, $g_s=0.8$; Table~\ref{tab:1}. }
  \label{fig:hairpin1}
\end{figure*}

We examined multi-agent swarming dynamics in the hairpin maze under several
conditions: pure swarming (\eq{oja}; \panel{hairpin1}{A}); swarming with sensory
cue inputs (\eq{netcue}; \panel{hairpin1}{B}); and swarming with sensory cue
inputs and reward approach (\eq{oja_r}; \panel{hairpin1}{C}). The sample
frames shown in \fig{hairpin1} demonstrate the emergence of phase-ordered
structures in each of these conditions with the clear distinction that tightly
configured reward rings became prevalent when reward learning was activated
\panelp{hairpin1}{c}. In that condition, with the same NeuroSwarms features as
studied above in the multireward arena, it was also clear that agents in the
second and third hallways had difficulty leaving to find another hallway with
a reward. We expected this was due to (1) the parity of swarming and reward
spatial constants ($\sigma$, $\kappa$), which perhaps overemphasized swarming
at the cost of reward-following in highly-partitioned environments, and (2) the
need for more sensitive reward-capture. Thus, we simulated a condition with
fixed and capturable rewards using $d_{\rm rad}=10$~points but also increased
the spatial constants with 3.3-fold bias for the reward value $(\sigma,
\kappa)=(2, 6.6)$ (\eqs{Ws}{Wr}; Table~\ref{tab:1}). Multi-agent trajectories
for this enhanced reward-exploration regime are shown in \panel{hairpin1}{d}:
with fixed rewards (top panel), the reward attractors dominate the dynamics and
agents generally stayed within their initial hallways; with capturable rewards
(bottom panel), there was substantially more path variability between agents,
spatial coverage increased (cf. the spiral patterns characteristic of agents'
exits from reward locations after contact), and many more agents were able to
traverse from one hallway to the next.

\begin{figure*}[tb]
  \includegraphics[width=1\textwidth]{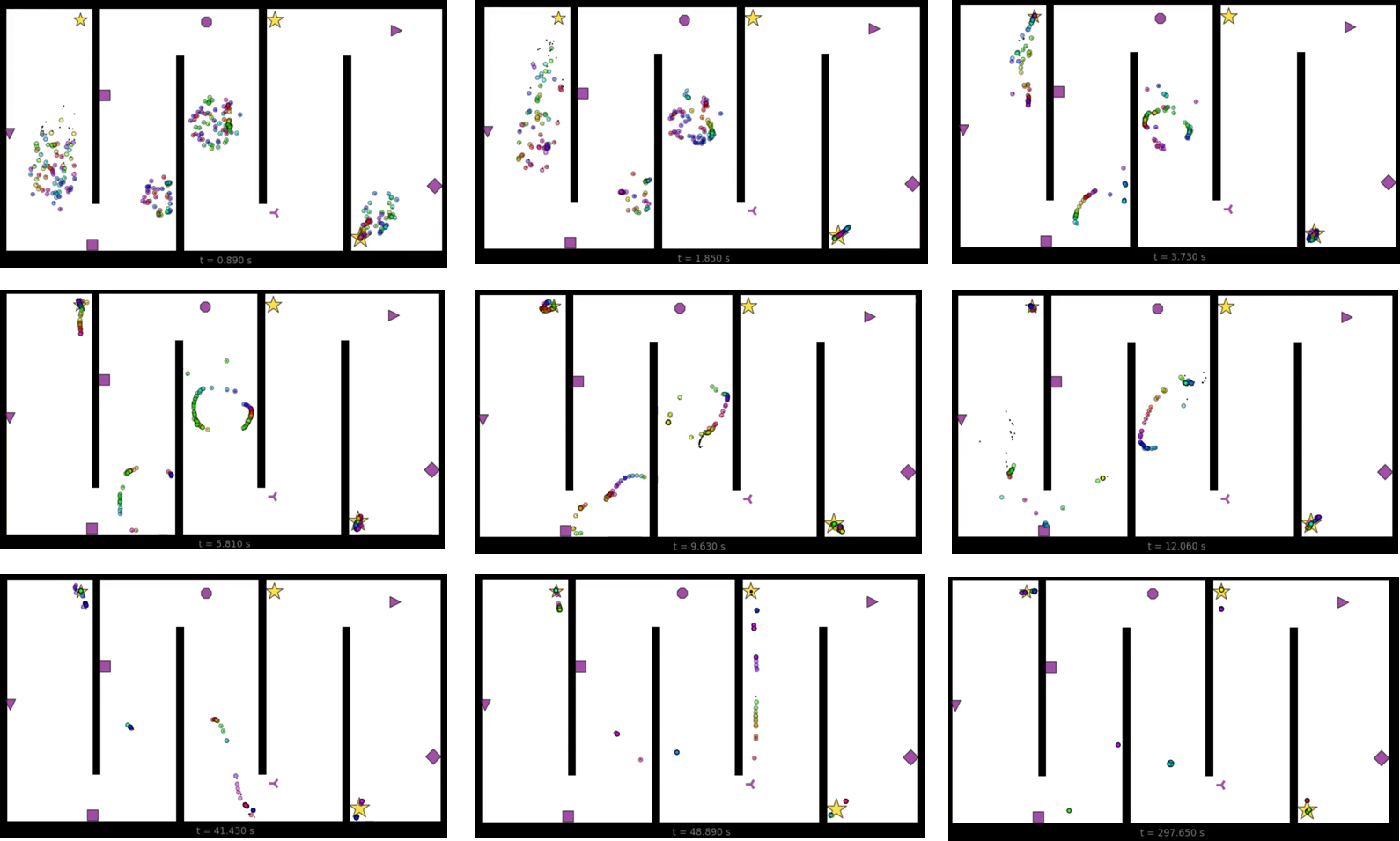}
  \caption{Multi-agent ring formations and trajectory sequences in the hairpin maze.
  Frames from a single simulation are shown for elapsed simulation times (s) from
  left-to-right, top-to-bottom: 0.89, 1.85, 3.73, 5.81, 9.63, 12.06, 41.43,
  48.99, 297.65.
  NeuroSwarms parameters: $\mathrm{duration}=300.0$, $d_{\rm max}=1.5$,
  $\sigma=2$, $\kappa=6.6$, $g_c=0.1$, $g_r=0.1$, $g_s=0.8$.}
  \label{fig:hairpin2}
\end{figure*}

To assess the converged state of multi-agent dynamics in the hairpin maze, we
simulated $N=300$ agents for 300~s using the same parameters and fixed rewards
as the top panel of \panel{hairpin1}{d}. The temporal progression of swarm state
across the simulation frames presented in \fig{hairpin2} shows distinct stages
exhibited by the four initial clusters of the swarm. The two clusters that
spawned in reward-free hallways eventually found their way around the barriers
to adjacent hallways after milling in various line segment or ring formations
for nearly a minute \figp{hairpin2}. All of the clusters successfully converged
onto the three reward locations in the maze, but the two that traversed hallways
left some agents behind. The progression of those swarm clusters from initial
positions to ring/arc formations to linear trajectory sequences to fixed-point
reward attractors illustrates a high degree of spontaneous adaptation to the
circumstances in the hairpin maze. These dynamics were self-organized and
emergent, providing behaviors that resulted in nearly complete convergence to
reward locations. Thus, NeuroSwarms demonstrated autonomous spatial navigation
to unknown, occluded, and remote rewards in a large and complex environment.

\section{Discussion} \label{sec:discussion}

We introduced the NeuroSwarms controller as a model for studying neural control
paradigms of artificial swarming agents. We demonstrated that NeuroSwarms
also acts as a two-way bridge between artificial systems and theoretical
models of animal cognition. This reciprocity arises due to a single-entity
paradigm in which NeuroSwarms controls a single agent in which an internal,
virtual `cognitive swarm' guides the agent's spatial behavior. Both modes of
operation, multi-agent and single-entity, share the same underlying neural
mechanisms (with differences described in Section~\ref{sec:singlemethods}).
This duality enables developments in artificial systems to also inform advances
in neurobiological theories of spatial cognition. Additionally, this duality
will aid discovery of neural dynamics in large, uncertain, and/or complex
environments based on closed-system approaches to distributed spatial coding. We
presented behaviors responding to environmental complexities such as multiple
reward sites (that optionally interact with the system by being `consumed' by
agents), heterogeneous agent-based preferences for neutral-valued spatial cues,
and geometric constraints that occlude agents' visibility of cues, rewards, and
other agents.

Swarms governed by NeuroSwarms self-organize into emergent, transitory
configurations in position and phase that directly recall spatial attractor
dynamics~\citep{Zhan96,Tsod99,SamsMcNa97,HedrZhan16,KnieZhan12} and sequential
oscillatory phenomena~\citep{OKeeRecc93,FostWils07,DrieTodo18,MonaDe-G19} that
have been theorized to operate within hippocampal circuits. We explicitly
designed NeuroSwarms to combine features of attractor maps and oscillatory
computing using robust transformations (such as the spatial kernels of distance
converted to synaptic strengths in \eqs{Ws}{Wr}). However, we did not anticipate
how readily such a system would self-organize into a variety of dynamic
spatiotemporal structures that recombined in complex patterns while supporting
navigation through our environments. A weakness of the presented implementation
was the use of a global, shared oscillation without allowing for noise, drift,
or independent perturbations~\citep[cf.][]{ZillHass10,MonaKnie11,MonaDe-G19}.
A more decentralized approach might utilize resonant agent-oscillators that
self-organize local oscillations depending on available information, task
requirements, or context. Such bottom-up oscillations might aggregate into a
global, swarm-wide oscillation under certain conditions, which should be studied
in future models.

To leverage inertial, energetic, and cost benefits of small-scale robots,
critical future applications of autonomous technologies may depend on
coordinating large numbers of agents with minimal onboard sensing and
communication resources. However, a critical problem for autonomous multi-robot
groups is that state-of-the-art control schemes break down as robotic agents
are scaled down (decreasing agent resources) and the numerical size of
swarms is scaled up (increasing communication and coordination requirements)
\citep{Murr07,HamaKhal16,YangBell18,ChunPara18}. NeuroSwarms addresses the
hypothesis that a similar distributed scaling problem may have been solved
by the evolved neural architecture of mammalian brains. Compared to signal
comprehension, signal production errors may be particularly deleterious to
large-scale, distributed and decentralized computations~\citep{SalaRouh19}.
Thus, onboard suites for future ‘cognitive swarming’ platforms based on
NeuroSwarms principles should emphasize reliable transmission of low-bandwidth
data packets (e.g., spikes or continuous phase signals). Low fidelity inputs are
more easily compensated by distributed processing; i.e., sensor designs should
emphasize energy and cost to maximize deployment duration and swarm size.

In summary, we made the explicit analogy from agents and swarms to neurons and
neural circuits. This analogy permitted the tools of theoretical neuroscience
to be leveraged in developing a model artificial spatial system. The
NeuroSwarms controller required two features to support cognitive swarming:
(1) an internal phase state, and (2) decoupling of physical location from
internal self-localization. The phase state naturally encapsulated neural
activation (cf. \eq{theta}) and could be used to drive spike generation,
if desired, in future models. Phase-based organization additionally
leveraged the expressive complexity of mobile oscillators revealed by the
swarmalator formalism~\citep{OKeeHong17,monaco2019cognitive}. The separation
of position vs.~self-localization allowed swarm motion dynamics to be
interpreted as Hebbian learning in an oscillatory place-coding neural network
(Section~\ref{sec:learning}). Thus, theorized hippocampal phenomena such as
attractor map formation and oscillatory sequence generation provide a framework
for advances in decentralized swarm control and, reciprocally, the theoretical
neuroscience of spatial navigation in complex, changing environments.

\begin{acknowledgements}
  The authors thank Marc Burlina and Marisel Villafane-Delgado for preliminary
  analyses and Robert Chalmers for helpful discussions about the manuscript.
\end{acknowledgements}

\section*{Conflict of interest}

The authors have no conflicts of interest to declare.

\bibliographystyle{spbasic}       
\bibliography{cognitive-swarming-minimal} 

\begin{thebibliography}{72}
\providecommand{\natexlab}[1]{#1}
\providecommand{\url}[1]{{#1}}
\providecommand{\urlprefix}{URL }
\expandafter\ifx\csname urlstyle\endcsname\relax
  \providecommand{\doi}[1]{DOI~\discretionary{}{}{}#1}\else
  \providecommand{\doi}{DOI~\discretionary{}{}{}\begingroup
  \urlstyle{rm}\Url}\fi
\providecommand{\eprint}[2][]{\url{#2}}

\bibitem[{Babichev and Dabaghian(2017)}]{BabiDaba17}
Babichev A, Dabaghian Y (2017) Transient cell assembly networks encode stable
  spatial memories. Scientific Reports 7(1):3959

\bibitem[{Bellmund et~al.(2018)Bellmund, G{\"a}rdenfors, Moser, and
  Doeller}]{BellGard18}
Bellmund JLS, G{\"a}rdenfors P, Moser EI, Doeller CF (2018) Navigating
  cognition: Spatial codes for human thinking. Science 362(6415)

\bibitem[{Burgess(2014)}]{Burg14}
Burgess N (2014) The 2014 nobel prize in physiology or medicine: A spatial
  model for cognitive neuroscience. Neuron 84(6):1120--1125

\bibitem[{Buzs{\'a}ki(1989)}]{Buzs89}
Buzs{\'a}ki G (1989) Two-stage model of memory trace formation: a role for
  ``noisy'' brain states. Neuroscience 31(3):551--70

\bibitem[{Buzs{\'a}ki(2005)}]{Buzs05}
Buzs{\'a}ki G (2005) Theta rhythm of navigation: link between path integration
  and landmark navigation, episodic and semantic memory. Hippocampus
  7(15):827--40

\bibitem[{Buzs{\'a}ki and Moser(2013)}]{BuzsMose13}
Buzs{\'a}ki G, Moser EI (2013) Memory, navigation and theta rhythm in the
  hippocampal-entorhinal system. Nat Neurosci 16(2):130

\bibitem[{Casali et~al.(2019)Casali, Bush, and Jeffery}]{CasaBush19}
Casali G, Bush D, Jeffery K (2019) Altered neural odometry in the vertical
  dimension. Proceedings of the National Academy of Sciences 116(10):4631--4636

\bibitem[{Chung et~al.(2018)Chung, Paranjape, Dames, Shen, and
  Kumar}]{ChunPara18}
Chung SJ, Paranjape AA, Dames P, Shen S, Kumar V (2018) A survey on aerial
  swarm robotics. IEEE Transactions on Robotics 34(4):837--855

\bibitem[{Climer et~al.(2013)Climer, Newman, and Hasselmo}]{ClimNewm13}
Climer JR, Newman EL, Hasselmo ME (2013) Phase coding by grid cells in
  unconstrained environments: two-dimensional phase precession. Eur J Neurosci
  38(4):2526--41

\bibitem[{Drieu et~al.(2018)Drieu, Todorova, and Zugaro}]{DrieTodo18}
Drieu C, Todorova R, Zugaro M (2018) Nested sequences of hippocampal assemblies
  during behavior support subsequent sleep replay. Science 362(6415):675--679

\bibitem[{Eichenbaum(2018)}]{Eich18}
Eichenbaum H (2018) Barlow versus {Hebb}: When is it time to abandon the notion
  of feature detectors and adopt the cell assembly as the unit of cognition?
  Neuroscience Letters 680:88--93

\bibitem[{Feng et~al.(2015)Feng, Silva, and Foster}]{FengSilv15}
Feng T, Silva D, Foster DJ (2015) Dissociation between the experience-dependent
  development of hippocampal theta sequences and single-trial phase precession.
  J Neurosci 35(12):4890--902

\bibitem[{Fenton et~al.(2008)Fenton, Kao, Neymotin, Olypher, Vayntrub, Lytton,
  and Ludvig}]{FentKao08}
Fenton AA, Kao HY, Neymotin SA, Olypher A, Vayntrub Y, Lytton WW, Ludvig N
  (2008) Unmasking the {CA}1 ensemble place code by exposures to small and
  large environments: more place cells and multiple, irregularly arranged, and
  expanded place fields in the larger space. J Neurosci 28(44):11250--62

\bibitem[{Foster(2017)}]{Fost17}
Foster DJ (2017) Replay comes of age. Annu Rev Neurosci 40:581--602

\bibitem[{Foster and Wilson(2007)}]{FostWils07}
Foster DJ, Wilson MA (2007) Hippocampal theta sequences. Hippocampus
  17(11):1093--1099

\bibitem[{Gaussier et~al.(2019)Gaussier, Banquet, Cuperlier, Quoy, Aubin,
  Jacob, Sargolini, Save, Krichmar, and Poucet}]{GausBanq19}
Gaussier P, Banquet JP, Cuperlier N, Quoy M, Aubin L, Jacob PY, Sargolini F,
  Save E, Krichmar JL, Poucet B (2019) Merging information in the entorhinal
  cortex: what can we learn from robotics experiments and modeling? Journal of
  Experimental Biology 222(Suppl 1)

\bibitem[{Gazi and Passino(2011)}]{gazi2011swarm}
Gazi V, Passino KM (2011) Swarm stability and optimization. Springer Science \&
  Business Media

\bibitem[{Gupta et~al.(2010)Gupta, van~der Meer, Touretzky, and
  Redish}]{GuptMeer10}
Gupta AS, van~der Meer MAA, Touretzky DS, Redish AD (2010) Hippocampal replay
  is not a simple function of experience. Neuron 65(5):695--705

\bibitem[{Hamann et~al.(2016)Hamann, Khaluf, Botev, Divband~Soorati, Ferrante,
  Kosak, Montanier, Mostaghim, Redpath, Timmis et~al.}]{HamaKhal16}
Hamann H, Khaluf Y, Botev J, Divband~Soorati M, Ferrante E, Kosak O, Montanier
  JM, Mostaghim S, Redpath R, Timmis J, et~al. (2016) Hybrid societies:
  challenges and perspectives in the design of collective behavior in
  self-organizing systems. Frontiers in Robotics and AI 3:14

\bibitem[{Hartley et~al.(2014)Hartley, Lever, Burgess, and
  O'Keefe}]{HartLeve14}
Hartley T, Lever C, Burgess N, O'Keefe J (2014) Space in the brain: how the
  hippocampal formation supports spatial cognition. Philos Trans R Soc Lond B
  Biol Sci 369(1635):20120510

\bibitem[{{Hasselmo}(2018)}]{Hass18}
{Hasselmo} ME (2018) A model of cortical cognitive function using hierarchical
  interactions of gating matrices in internal agents coding relational
  representations. arXiv e-prints p arXiv:1809.08203

\bibitem[{Hebb(1949)}]{Hebb49}
Hebb DO (1949) The Organization of Behavior: A Neuropsychological Theory. Wiley
  and Sons, New York

\bibitem[{Hedrick and Zhang(2016)}]{HedrZhan16}
Hedrick KR, Zhang K (2016) Megamap: flexible representation of a large space
  embedded with nonspatial information by a hippocampal attractor network.
  Journal of Neurophysiology 116(2):868--891

\bibitem[{Jayakumar et~al.(2019)Jayakumar, Madhav, Savelli, Blair, Cowan, and
  Knierim}]{JayaMadh19}
Jayakumar RP, Madhav MS, Savelli F, Blair HT, Cowan NJ, Knierim JJ (2019)
  Recalibration of path integration in hippocampal place cells. Nature
  566(7745):533--537

\bibitem[{Jeewajee et~al.(2014)Jeewajee, Barry, Douchamps, Manson, Lever, and
  Burgess}]{JeewBarr14}
Jeewajee A, Barry C, Douchamps V, Manson D, Lever C, Burgess N (2014) Theta
  phase precession of grid and place cell firing in open environments. Philos
  Trans R Soc Lond B Biol Sci 369(1635):20120532

\bibitem[{Jensen and Lisman(2000)}]{JensLism00}
Jensen O, Lisman JE (2000) Position reconstruction from an ensemble of
  hippocampal place cells: contribution of theta phase coding. J Neurophysiol
  83(5):2602--9

\bibitem[{Johnson and Redish(2007)}]{JohnRedi07}
Johnson A, Redish AD (2007) Neural ensembles in {CA}3 transiently encode paths
  forward of the animal at a decision point. J Neurosci 27(45):12176--89

\bibitem[{Knierim(2006)}]{Knie06}
Knierim JJ (2006) Neural representations of location outside the hippocampus.
  Learn Mem 13(4):405--415

\bibitem[{Knierim and Hamilton(2011)}]{KnieHami11}
Knierim JJ, Hamilton DA (2011) Framing spatial cognition: Neural
  representations of proximal and distal frames of reference and their roles in
  navigation. Physiological Reviews 91(4):1245--1279

\bibitem[{Knierim and Zhang(2012)}]{KnieZhan12}
Knierim JJ, Zhang K (2012) Attractor dynamics of spatially correlated neural
  activity in the limbic system. Annu Rev Neurosci 35:267--85

\bibitem[{Kunz et~al.(2019)Kunz, Maidenbaum, Chen, Wang, Jacobs, and
  Axmacher}]{KunzMaid19}
Kunz L, Maidenbaum S, Chen D, Wang L, Jacobs J, Axmacher N (2019) Mesoscopic
  neural representations in spatial navigation. Trends in Cognitive Sciences

\bibitem[{Levy and Steward(1979)}]{LevyStew79}
Levy WB, Steward O (1979) Synapses as associative memory elements in the
  hippocampal formation. Brain Res 175(2):233--45

\bibitem[{Momennejad et~al.(2018)Momennejad, Otto, Daw, and
  Norman}]{MomeOtto18}
Momennejad I, Otto AR, Daw ND, Norman KA (2018) Offline replay supports
  planning in human reinforcement learning. eLife 7:e32548

\bibitem[{Monaco and Abbott(2011)}]{MonaAbbo11}
Monaco JD, Abbott LF (2011) Modular realignment of entorhinal grid cell
  activity as a basis for hippocampal remapping. J Neurosci 31(25):9414--25

\bibitem[{Monaco et~al.(2011)Monaco, Knierim, and Zhang}]{MonaKnie11}
Monaco JD, Knierim JJ, Zhang K (2011) Sensory feedback, error correction, and
  remapping in a multiple oscillator model of place-cell activity. Front Comput
  Neurosci 5:39

\bibitem[{Monaco et~al.(2014)Monaco, Rao, Roth, and Knierim}]{MonaRao14}
Monaco JD, Rao G, Roth ED, Knierim JJ (2014) Attentive scanning behavior drives
  one-trial potentiation of hippocampal place fields. Nat Neurosci
  17(5):725--731

\bibitem[{Monaco et~al.(2019{\natexlab{a}})Monaco, De~Guzman, Blair, and
  Zhang}]{MonaDe-G19}
Monaco JD, De~Guzman RM, Blair HT, Zhang K (2019{\natexlab{a}}) Spatial
  synchronization codes from coupled rate-phase neurons. PLoS Comput Biol
  15(1):e1006741

\bibitem[{Monaco et~al.(2019{\natexlab{b}})Monaco, Hwang, Schultz, and
  Zhang}]{monaco2019cognitive}
Monaco JD, Hwang GM, Schultz KM, Zhang K (2019{\natexlab{b}}) Cognitive
  swarming: an approach from the theoretical neuroscience of hippocampal
  function. In: Micro-and Nanotechnology Sensors, Systems, and Applications XI,
  International Society for Optics and Photonics, vol 10982, p 109822D

\bibitem[{Moser and Paulsen(2001)}]{MosePaul01}
Moser EI, Paulsen O (2001) New excitement in cognitive space: between place
  cells and spatial memory. Curr Opin Neurobiol 11(6):745--751

\bibitem[{Moser et~al.(2008)Moser, Kropff, and Moser}]{MoseKrop08}
Moser EI, Kropff E, Moser MB (2008) Place cells, grid cells, and the brain's
  spatial representation system. Annu Rev Neurosci 31(1):69--89

\bibitem[{Muessig et~al.(2019)Muessig, Lasek, Varsavsky, Cacucci, and
  Wills}]{MuesLase19}
Muessig L, Lasek M, Varsavsky I, Cacucci F, Wills TJ (2019) Coordinated
  emergence of hippocampal replay and theta sequences during post-natal
  development. Current Biology 29(5):834--840.e4

\bibitem[{Murray(2007)}]{Murr07}
Murray RM (2007) Recent research in cooperative control of multivehicle
  systems. Journal of Dynamic Systems, Measurement, and Control 129(5):571--583

\bibitem[{Ocko et~al.(2018)Ocko, Hardcastle, Giocomo, and Ganguli}]{OckoHard18}
Ocko SA, Hardcastle K, Giocomo LM, Ganguli S (2018) Emergent elasticity in the
  neural code for space. Proc Natl Acad Sci U S A 115(50):E11798--E11806

\bibitem[{Oja(1982)}]{Oja82}
Oja E (1982) Simplified neuron model as a principal component analyzer. Journal
  of Mathematical Biology 15(3):267--273

\bibitem[{O'Keefe and Dostrovsky(1971)}]{OKeeDost71}
O'Keefe J, Dostrovsky J (1971) The hippocampus as a spatial map: preliminary
  evidence from unit activity in the freely-moving rat. Brain Res
  34(1):171--175

\bibitem[{O'Keefe and Nadel(1978)}]{OKeeNade78}
O'Keefe J, Nadel L (1978) The {H}ippocampus as a {C}ognitive {M}ap. Clarendon
  Press, Oxford, UK

\bibitem[{{O'Keefe} and Recce(1993)}]{OKeeRecc93}
{O'Keefe} J, Recce ML (1993) Phase relationship between hippocampal place units
  and the {EEG} theta rhythm. Hippocampus 3(3):317--30

\bibitem[{O'Keeffe et~al.(2017)O'Keeffe, Hong, and Strogatz}]{OKeeHong17}
O'Keeffe KP, Hong H, Strogatz SH (2017) Oscillators that sync and swarm. Nature
  communications 8(1):1504

\bibitem[{{\'O}lafsd{\'o}ttir et~al.(2018){\'O}lafsd{\'o}ttir, Bush, and
  Barry}]{OlafBush18}
{\'O}lafsd{\'o}ttir HF, Bush D, Barry C (2018) The role of hippocampal replay
  in memory and planning. Curr Biol 28(1):R37--R50

\bibitem[{Papale et~al.(2016)Papale, Zielinski, Frank, Jadhav, and
  Redish}]{PapaZiel16}
Papale AE, Zielinski MC, Frank LM, Jadhav SP, Redish AD (2016) Interplay
  between hippocampal sharp-wave-ripple events and vicarious trial and error
  behaviors in decision making. Neuron 92(5):975--982

\bibitem[{Pfeiffer and Foster(2013)}]{PfeiFost13}
Pfeiffer BE, Foster DJ (2013) Hippocampal place-cell sequences depict future
  paths to remembered goals. Nature 497(7447):74--9

\bibitem[{Poll et~al.(2016)Poll, Nguyen, and Kilpatrick}]{PollNguy16}
Poll DB, Nguyen K, Kilpatrick ZP (2016) Sensory feedback in a bump attractor
  model of path integration. J Comput Neurosci 40(2):137--55

\bibitem[{Poulter et~al.(2018)Poulter, Hartley, and Lever}]{PoulHart18}
Poulter S, Hartley T, Lever C (2018) The neurobiology of mammalian navigation.
  Current Biology 28(17):R1023--R1042

\bibitem[{Renn{\'o}-Costa and Tort(2017)}]{RennTort17}
Renn{\'o}-Costa C, Tort ABL (2017) Place and grid cells in a loop: implications
  for memory function and spatial coding. J Neurosci 37(34):8062--8076

\bibitem[{Rich et~al.(2014)Rich, Liaw, and Lee}]{RichLiaw14}
Rich PD, Liaw HP, Lee AK (2014) Large environments reveal the statistical
  structure governing hippocampal representations. Science 345(6198):814--817

\bibitem[{Salahshour et~al.(2019)Salahshour, Rouhani, and Roudi}]{SalaRouh19}
Salahshour M, Rouhani S, Roudi Y (2019) Phase transitions and asymmetry between
  signal comprehension and production in biological communication. Scientific
  Reports 9(1):3428

\bibitem[{Samsonovich and McNaughton(1997)}]{SamsMcNa97}
Samsonovich A, McNaughton BL (1997) Path integration and cognitive mapping in a
  continuous attractor neural network model. J Neurosci 17(15):5900--5920

\bibitem[{Savelli and Knierim(2010)}]{SaveKnie10}
Savelli F, Knierim JJ (2010) Hebbian analysis of the transformation of medial
  entorhinal grid-cell inputs to hippocampal place fields. J Neurophysiol
  103(6):3167--83

\bibitem[{Savelli and Knierim(2019)}]{SaveKnie19}
Savelli F, Knierim JJ (2019) Origin and role of path integration in the
  cognitive representations of the hippocampus: computational insights into
  open questions. J Exp Biol 222(Pt Suppl 1)

\bibitem[{Savelli et~al.(2008)Savelli, Yoganarasimha, and Knierim}]{SaveYoga08}
Savelli F, Yoganarasimha D, Knierim JJ (2008) Influence of boundary removal on
  the spatial representations of the medial entorhinal cortex. Hippocampus
  18(12):1270--1282

\bibitem[{Schiller et~al.(2015)Schiller, Eichenbaum, Buffalo, Davachi, Foster,
  Leutgeb, and Ranganath}]{SchiEich15}
Schiller D, Eichenbaum H, Buffalo EA, Davachi L, Foster DJ, Leutgeb S,
  Ranganath C (2015) Memory and space: Towards an understanding of the
  cognitive map. J Neurosci 35(41):13904--13911

\bibitem[{Tomov et~al.(2018)Tomov, Yagati, Kumar, Yang, and
  Gershman}]{TomoYaga18}
Tomov M, Yagati S, Kumar A, Yang W, Gershman S (2018) Discovery of hierarchical
  representations for efficient planning. bioRxiv 499418

\bibitem[{Tsodyks(1999)}]{Tsod99}
Tsodyks M (1999) Attractor neural network models of spatial maps in
  hippocampus. Hippocampus 9(4):481--9

\bibitem[{Vanderwolf(1969)}]{Vand69}
Vanderwolf CH (1969) Hippocampal electrical activity and voluntary movement in
  the rat. Electroencephalogr Clin Neurophysiol 26(4):407--18

\bibitem[{Wang et~al.(2018)Wang, Chen, Lee, Deshmukh, Yoganarasimha, Savelli,
  and Knierim}]{WangChen18}
Wang C, Chen X, Lee H, Deshmukh SS, Yoganarasimha D, Savelli F, Knierim JJ
  (2018) Egocentric coding of external items in the lateral entorhinal cortex.
  Science 362(6417):945--9

\bibitem[{Wang et~al.(2016)Wang, Roth, and Pastalkova}]{WangRoth16}
Wang Y, Roth Z, Pastalkova E (2016) Synchronized excitability in a network
  enables generation of internal neuronal sequences. Elife 5

\bibitem[{Wikenheiser and Redish(2015)}]{WikeRedi15}
Wikenheiser AM, Redish AD (2015) Hippocampal theta sequences reflect current
  goals. Nat Neurosci 18:289--294

\bibitem[{Yadav and Doreswamy(2017)}]{YadaDore17}
Yadav CK, Doreswamy Y (2017) Scale invariance in lateral head scans during
  spatial exploration. Physical Review Letters 118(15):158104

\bibitem[{Yang et~al.(2018)Yang, Bellingham, Dupont, Fischer, Floridi, Full,
  Jacobstein, Kumar, McNutt, Merrifield et~al.}]{YangBell18}
Yang GZ, Bellingham J, Dupont PE, Fischer P, Floridi L, Full R, Jacobstein N,
  Kumar V, McNutt M, Merrifield R, et~al. (2018) The grand challenges of
  {Science Robotics}. Science Robotics 3(14):eaar7650

\bibitem[{Yartsev and Ulanovsky(2013)}]{YartUlan13}
Yartsev MM, Ulanovsky N (2013) Representation of three-dimensional space in the
  hippocampus of flying bats. Science 340(6130):367--72

\bibitem[{Zhang(1996)}]{Zhan96}
Zhang K (1996) Representation of spatial orientation by the intrinsic dynamics
  of the head-direction cell ensemble: a theory. J Neurosci 16(6):2112--26

\bibitem[{Zilli and Hasselmo(2010)}]{ZillHass10}
Zilli EA, Hasselmo ME (2010) Coupled noisy spiking neurons as
  velocity-controlled oscillators in a model of grid cell spatial firing. J
  Neurosci 30(41):13850--60

\end{thebibliography}

\end{document}